\documentclass[10pt,a4paper,onecolumn]{article}

\usepackage[journal=jpclcd]{achemso}
\setkeys{acs}{articletitle = true}
\usepackage[T1]{fontenc}       
\usepackage{graphicx}
\usepackage{dblfloatfix}
\usepackage{placeins}
\usepackage{float}
\usepackage{authblk}
\usepackage{a4wide}
\usepackage{setspace}
\usepackage[margin=1in]{geometry}
\usepackage{physics}
\usepackage{color}
\usepackage{bm}
\usepackage[normalem]{ulem}
\usepackage{booktabs, tabularx} 
\newcolumntype{C}{>{\centering\arraybackslash}X}

\author[1]{Asha S. Thomas}
\author[1]{Camelia Roy}
\author[1]{Vivek N. Bhat}
\author[1]{Indranil Roy}
\author[1]{Vivek Tiwari \thanks{vivektiwari@iisc.ac.in}}
\affil[1]{Solid State and Structural Chemistry Unit, Indian Institute of Science, Bangalore, Karnataka 560012, India}

\newcolumntype{C}{>{\centering\arraybackslash}X}

\newcommand*{\vt}[1]{\textcolor{black}{ #1}}
\newcommand*{\si}[1]{\textcolor{black}{ #1}}

\begin{document}

\title{Disordered Photosynthetic Aggregates Can Host Functional Vibronic Couplings At Room Temperature}
\maketitle
	
\begin{abstract}
Photosynthesis relies on a network of chlorophyll-like molecules which together lead to efficient long-range energy funneling. Evidence at cryogenic temperatures suggests that mechanistic details of energy/charge transfer must invoke delocalized vibronic states. Whether these survive at physiological temperature in large photosynthetic aggregates is an open question. Parallel research on artificial templates has relied on cyanines which are unlike chlorophylls. We report two-dimensional electronic spectra of porphyrin nanotubes where we selectively probe mixed $Q_x-Q_y$ states through polarization control.  Early time cross-peaks, their rapid broadening and survival of anisotropic $Q_x-Q_y$ quantum beats conclusively demonstrate that overlapping vibrational-electronic bands of photosynthetic aggregates indeed host functional vibronic couplings at room temperature. Calculations reveal that disorder is the vital ingredient that dramatically enhances $Q_x-Q_y$ vibronic mixing across the entire $Q$ band.  The parameter regime where energetic disorder is of the order of dense Raman-active vibrations with weak reorganization energies may be the key design principle. \\

\end{abstract}

\section*{Introduction} \label{section1}
Internal conversion through overlapping vibrational-electronic manifolds\cite{Robinson1962} dictates the fate of photoexcitations in natural and artificial light harvesting. In case of natural photosynthesis\cite{blankenship2021molecular}, delocalized photoexcitations created in pigment-protein complexes (antennas) survive lossy internal conversion channels up to several hundred picosecond timescales, diffuse across tens of nanometer (nm) lengthscales across a network of such antenna proteins to ultimately be harvested with near unity quantum efficiency as a stabilized charge-separated state . Cryogenic experiments and theoretical proposals have suggested a functional role for non-classical vibrations\cite{Makri2022, OlayaCastro2014} which can promote mixed vibrational-electronic (vibronic) states and nested energy funnels within these proteins. Whether such mechanistic principles survive decoherence at room temperature in large aggregates remains a hotly debated question\cite{Miller2020, scholes2017using}, inspiring several design strategies\cite{Brixner2017a, scholes2017using} to understand and mimic photosynthetic energy transfer. Probing this question in the context of large photosynthetic aggregates at room temperature is the subject of this manuscript. \\

Artificial light harvesting templates aiming to mimic the natural design principles have largely included covalently linked dimers\cite{Bocian2023}, self-assembled nanotubes\cite{Caram2016} and DNA templates hosting molecular networks with precise control\cite{Cohen2021} over position, orientation and the resulting excitonic couplings. Among these, self-assembled light harvesting nanotubes\cite{Kohler2024} (LHNs), specifically cyanine LHNs, represent a distinctly different approach where energy funneling across nanotube layers\cite{Milota2010,Kriete2017}, robust excitons with estimated coherence length over 1000 molecules and extraordinary micron-scale exciton diffusion lengths\cite{Caram2016} have been demonstrated.\\

\begin{figure}
\centering
\includegraphics[width=4 in]{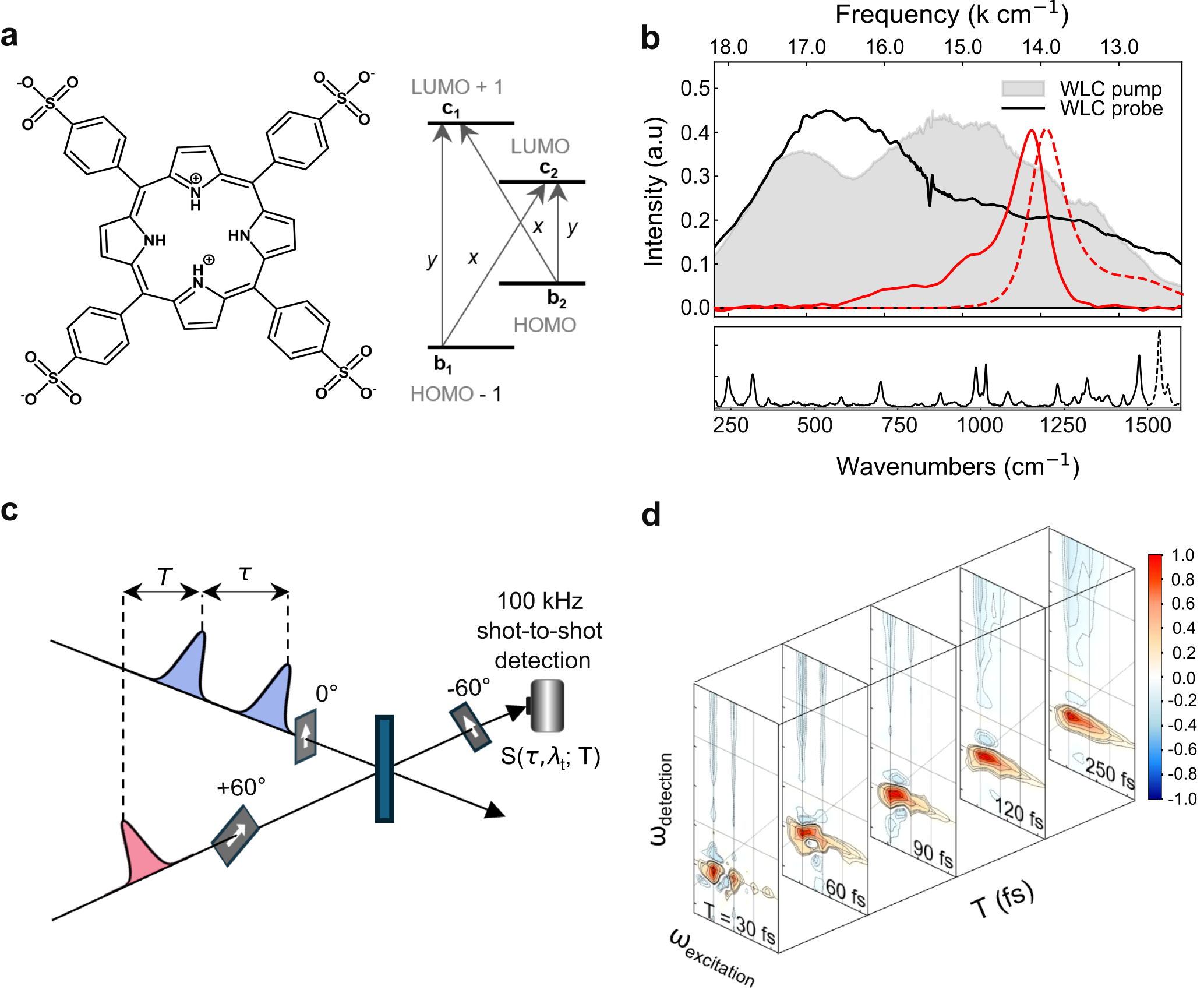}
    \caption{\footnotesize\textbf{ Porphyrin nanotubes share vibronic features with photosynthetic light harvesting systems.} \textbf{a.} Molecular and electronic structure of the constituent tetra(4-sulfonatophenyl) porphyrin (TPPS) derivative at pH 4. Electrostatic interactions between the positive core and the negative sulfonato groups allows for formation of large nanotube aggregates. Akin to chlorophylls, the four frontier orbitals describe\cite{gouterman1961spectra} the $x$ and $y$ polarized transitions in the $Q$ and $B$ bands. \textbf{b.} (top) $Q$ band absorption (solid) spectrum and emission (dashed) spectrum of nanotube aggregate overlaid with the broadband pump and probe spectra. (bottom) Raman spectrum of porphyrin nanotubes excited at 532 nm. Peaks after 1500 cm$^{-1}$ have been scaled by 0.5$\times$. \textbf{c.} Partially-collinear 2DES experiment\cite{thomas2023rapid} with polarization control on pump and probe arms and 100 kHz shot-to-shot detection. The additional knob of polarization control allows for selecting mixed $Q_x-Q_y$ states.\textbf{d.} 2D contour maps decongest overlapping vibronic bands by correlating detection and excitation energy of a system as a function of pump-probe waiting times $T$. }
\label{figure1}
\end{figure}

An interesting feature that differentiates all above approaches for designing artificial light-harvesting antennas from the naturally occurring photosynthetic nanotubes and bundles (chlorosomes\cite{oostergetel2010chlorosome}) is that the constituent molecules of the latter rely on chlorophyll (Chl) or closely related molecules that share a unique electronic structure with $Q_x$ and $Q_y$ bands and $B_x$ and $B_y$ bands in the visible and ultraviolet regions, respectively. Here the subscripts \textit{x} and \textit{y} denote directions of electronic transition dipole moments. Such an electronic structure is well known to be sensitive to substituents and coordination of ring nitrogens\cite{gouterman1961spectra}, undergoes fast 100-200 fs internal conversion\cite{Ogilvie2019} between $Q_x$ and $Q_y$ bands as well as between the $B$ and $Q$ bands\cite{Joo2015}, and hosts a dense spectrum low-frequency Raman active vibrations\cite{diers1995qy} that undergo negligible Franck-Condon (FC) displacements (dimensionless displacement d $<<$ 1) upon photoexcitation\cite{ratsep2011demonstration}. Small FC displacements maximize the non-adiabatic coupling between vibrational and electronic wavefunctions thus creating nested energy funnels\cite{tiwari2013electronic,Jonas2018} in several photosynthetic proteins\cite{Jonas2018}. In contrast, in case of large nuclear displacements, non-adiabatic vibronic mixing between the donor-acceptor excitons only occurs in a limited range of nuclear coordinates\cite{tiwari2014vibronic} thereby limiting the efficacy of nested energy funnels. Thus artificial dimers, nanotubes, or DNA templates based on cyanine derivatives likely operate in parameter regimes distinctly different from natural photosynthesis\cite{tiwari2014vibronic}. This lack of similarity between artificial aggregates and natural photosynthetic antennas motivates study of those artificial LHN\cite{Knoester2019} templates where the mechanistic details of photosynthetic energy and charge transfer can be reproduced and understood. The schematic in \si{Figure~\ref{figure1}a} illustrates the similarity in the four frontier orbital electronic structure\cite{gouterman1961spectra} of the constituent porphyrin molecules employed in this work with that known in chloropylls. Similar to the 100-200 fs internal conversion known\cite{Hauer2025,Ogilvie2019} in chlorophylls, sub-100 fs internal conversion\cite{Joo2015,Mauiri2024}, between $B-Q$ and $Q_x-Q_y$ bands is known in porphyrins as well. \si{Figure~\ref{figure1}b} shows the $Q$ bands of the porphyrin nanotube aggregate investigated here. The theoretical predictions\cite{Stradomska2010b} of the current vibronic exciton models for nanotube aggregates formed from these constituent porphyrins suggest that the main band is $Q_x$ in character and polarized along the cylindrical axis while the first shoulder is $Q_y$ in character and polarized perpendicular to the cylindrical axis.  The lower panel of \si{Figure~\ref{figure1}b} shows the dense Raman active vibrational spectrum\cite{Akins1994} of the porphyrin nanotubes which closely mimics those of photosynthetic pigments. Similar to energy funnels in photosynthetic proteins\cite{Jonas2018}, the resulting overlapping vibrational and electronic bands suggest that vibrational excitations in the lower electronic state may lead to near-resonant vibronic mixing with the upper electronic states within the main $Q$ band as well as between the main $Q$ band and its first shoulder. The absorption and Raman spectra (\si{Figure~\ref{figure1}b}) together illustrate the central question that this work aims to address, that is, whether such near-resonant vibronic mixing effects survive at room temperature to play a functional role, that is, drive internal conversion in large, disordered photosynthetic aggregates ? \\ 

Two-dimensional electronic spectroscopy\cite{jonas2003two} (2DES) has provided unprecedented insights\cite{Jonas2018, Beck2022} into the dynamics that emerge from overlapping vibrational and electronic bands due to its fast temporal resolution and spectral decongestion along excitation, detection and coherence axes. As shown in \si{Figures~\ref{figure1}c,d}, in our  partially-collinear pump-probe\cite{OgilvieARPC} 2DES experiment, a pair of pump pulses create absorptive changes in the sample as a function of their mutual delay $\tau$. A delayed probe pulse measures these changes as a function of pump-probe waiting time $T$ and the probe detection wavelength $\lambda_t$ resulting in the differential probe spectrum $S (\tau, T, \lambda_t)$. Division by the probe spectrum $S(\lambda)$ with no pump, Fourier transformation along $\tau$ and Jacobian conversion to $\omega_t$ leads to 2D contour maps that correlate detection and excitation frequencies of a given system as a function of $T$. Extensions of concepts from electronic polarization anisotropy to 2DES have also led to polarization-controlled 2DES (P-2D) \cite{read2008visualization} where 2D cross-peaks\cite{farrell2022polarization,sohoni2023phycobilisome}, as well as vibronically enhanced quantum beats\cite{palevcek2017quantum,bhattacharyya2023low,sahu2025isolating}, can be uniquely isolated. \\

To best of our knowledge, all demonstrations\cite{Jonas2018} of P-2D so far in uniquely isolating vibronically mixed states in photosynthetic proteins have been conducted at cryogenic temperatures on small dimer-like aggregates. Here we present P-2D experiments on the $Q$ band of self-assembled porphyrin nanotubes revealing the rich complexity of overlapping vibronic bands that emerge in large self-assembled and disordered aggregates.  Our observations demonstrate that the overlapping vibronic bands in porphyrin nanotubes closely mimic all dynamical features of naturally occurring chlorosome nanotubes. These observations, that are unique to P-2D and not evident in any prior pump-probe experiments\cite{Kano2001,Wan2014}, include features such as 2D cross peaks (CPs) at $T$ as early as 30 fs, their broadening within 150 fs indicating rapid exciton delocalization, and anisotropic vibrational quantum beats that are unique\cite{sahu2025isolating,bhattacharyya2023low} reporters of $Q_x-Q_y$ mixed states arising from near-resonant vibronic couplings. This is contrary to what is predicted\cite{Stradomska2010b} by current vibronic exciton models for porphyrin nanotubes. Our refined vibronic exciton model for the nanotube suggests that disorder is \textit{the} vital ingredient that promotes $Q_x-Q_y$ electronic mixing across the entire $Q$ band. Given this unavoidable electronic mixing, intramolecular vibrations of porphyrins that tune energy gaps ensure that the energetically distinct $Q$-band and its overlapping vibronic shoulders are strongly mixed. The evidence for intraband vibronic couplings surviving in large aggregates at room temperature suggests that the mechanisms underlying internal conversion in photosynthesis may be surprisingly robust and disorder induced vibronic exciton delocalization may be a vital part of the photosynthetic design. \\

\section*{Results and Discussion} \label{section2}

\subsection*{2DES on porphyrin nanotubes reveals intra-Q band cross peaks} 

The 2DES experiment generates a non-linear third-order polarization which then radiates a 90$^o$ phase-shifted signal field. In the non-linear response function formalism, the four light-matter interactions are treated perturbatively and each term in the perturbative expansion of the response function can be represented\cite{MukamelBook} by a corresponding diagrammatic representation (wavemixing diagrams). The matter part of the non-linear response functions can be represented \cite{jonas2003two} as $R \sim \left\langle (\hat{\mu}.\hat{E}_a) (\hat{\mu}.\hat{E}_b) (\hat{\mu}.\hat{E}_c) (\hat{\mu}.\hat{E}_s)\right\rangle$ where the field polarization vectors $\hat{E}_{a-c}$ in the lab frame interact with the system to generate the radiated signal field $\hat{E}_{s}$, which is then detected by interference with a known field, in our case $\vec{E}_c$. $\hat{\mu}$ represent the transition dipole vectors in the molecular frame between the initial and final system eigenstates involved at each interaction. $\left\langle ..\right\rangle$ denotes orientational averaging between the molecular and lab frames. \\ 

\begin{figure}
\centering
\includegraphics[width=5 in]{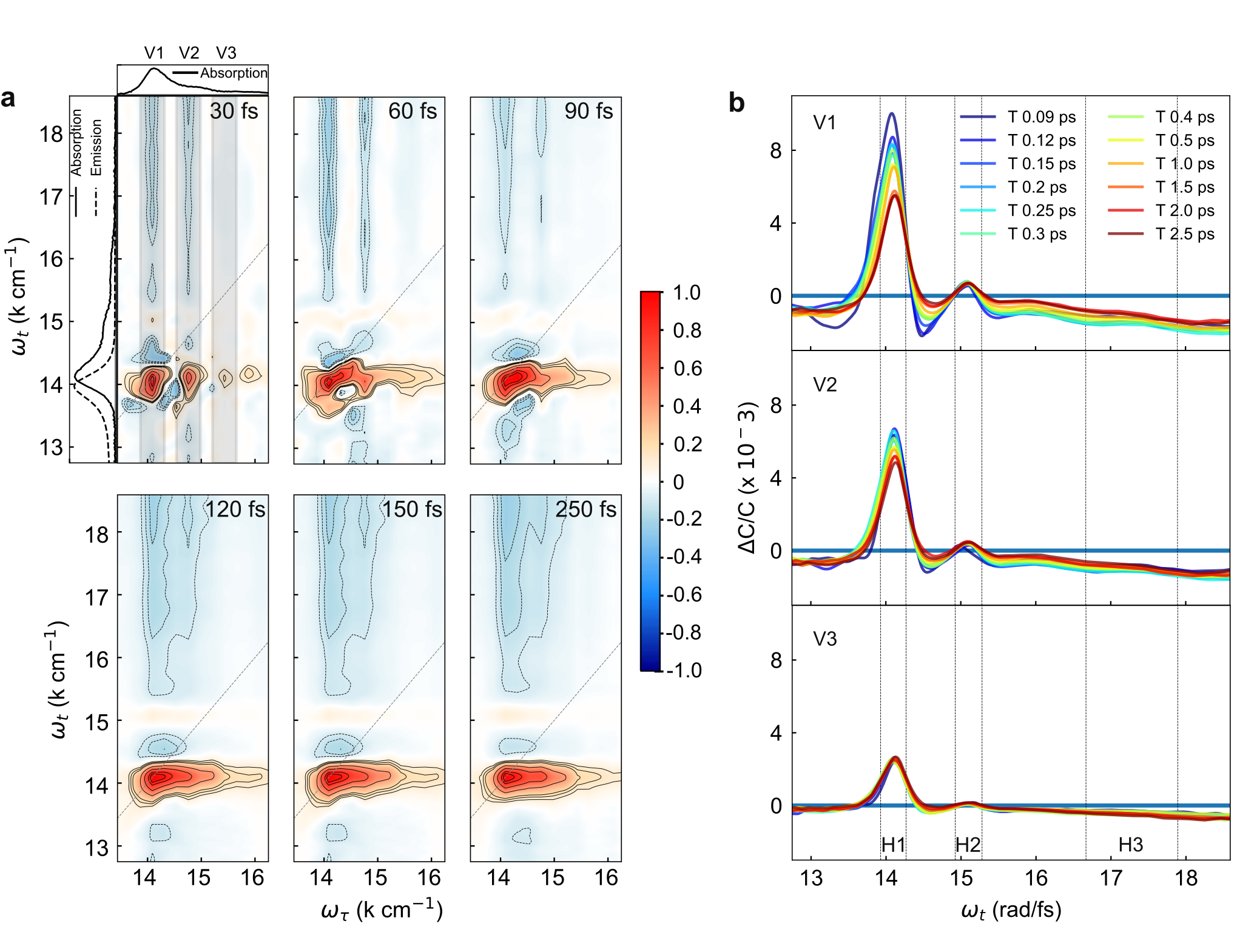}
\caption{\footnotesize \textbf{Rapid intra-$Q$ band delocalization revealed by early $T$ 2D spectra.} \textbf{a.} Normalized absorptive PE-2D spectra of porphyrin nanotube at a few representative \textit{T} points. Three vertical bands V1- V3 are centered on the 2D peaks in the CP$_L$ region seen in $T=30 fs$ 2D spectrum. \textbf{b.} Signal integrated along the vertical bands V1-V3 as a function of $T$. Each band is integrated across an energy band of 223.51 cm$^{-1}$. The vertical lines show the horizontal bands H1-H3. All panels are shown on a common y-scale such that changes with $T$ can be directly compared across panels. Contours are drawn at 10\%, 15\% and 20\% and then in steps of 20\%.}\label{figure2}
\end{figure}

\si{Figure~\ref{figure2}a} shows a series of PE-2D spectra of porphyrin nanotubes with orthogonal relative polarizations of the pump and probe electric fields. For waiting time \textit{T} as early as 30 fs, we observe two well-resolved CPs between the main $Q$ band ($Q_x$) and its two smaller vibronic shoulders ($Q_y$) seen in \si{Figure~\ref{figure1}d}. The two lower CPs (CP$_L$) in the V2 and V3 bands, each correspond to the two vibronic shoulders in the absorption spectrum, and suggest that $Q_y$ polarized states may be strongly coupled to the $Q_x$ polarized main $Q$ band. Interestingly, each absorption band and the corresponding CP is also accompanied by a distinct ESA band suggesting distinct transitions to the 2-quantum electronic manifold. The weak DPs corresponding to the vibronic shoulders in the absorption spectrum as well as weak upper CPs (CP$_U$, seen in the V1 band) are overwhelmed by the ESA bands. \\

Note that shared ground state correlations between coherently coupled molecules predicts ground state bleach (GSB) type upper and lower CPs between the $Q_x$ and $Q_y$ bands of the nanotube at $T$ = 0 fs, where bleach of one transition affects the other. GSB type peaks are expected to decay slowly. A weak CP$_U$ does indeed persist and seen clearly in the integrated V1 area in \si{Figure~\ref{figure2}b}. In comparison, the dynamics in the CP$_L$ region, main band in the integrated V2 area in \si{Figure~\ref{figure2}b}, is dominated by the excited state emission (ESE) signal. This is evident from $\sim$10$\times$ imbalanced CP$_L$ amplitude compared to CP$_U$ as early as $T=$ 30 fs, its significant decay on a 238 fs timescale (\textit{vide infra}) and rapid peak broadening in the CP$_L$ region which likely overwhelmes any GSB type CPs. Significant broadening of these features, including the distinct ESA bands, within 250 fs supports the idea of strong intraband couplings causing rapid exciton delocalization. Similar rapid broadening of 2D peakshapes has been observed in chlorosomes from green sulfur bacteria \cite{Zigmantas2012} and attributed to rapid intraband exciton delocalization. The resulting ultrafast loss of frequency-frequency correlations is reported directly by the 2D lineshapes. It is interesting to note that compared to cyanine nanotubes, both systems, the porphyrin nanotubes studied here and chlorosomes studied previously\cite{Zigmantas2012}, display significantly faster exciton delocalization and peak broadening (compare \si{Figure~\ref{figure2}} of current work and ref. \cite{Zigmantas2012} with Figure 5 and Figure 2 of ref.\cite{Milota2010} and ref.\cite{Kriete2019}, respectively). \\

\si{Figure~\ref{figure2}b} also analyzes the excitation frequency dependence of the decay kinetics in the three vertical bands V1-V3 marked in the $T$ = 30 fs 2D spectrum in panel a. The $T$ kinetic fits are shown in \si{Figure S7}. In V1, the lower diagonal peak (DP$_L$) undergoes a rapid decay along with a 2.9 nm blue shift within 2.5 ps. In contrast, CP$_U$ shows a slower \textit{T} decay with almost non-decaying amplitude in the 90 fs - 2.5 ps time window. In V2, the main band is CP$_L$, and compared to DP$_L$, it shows a slower decay and a less prominent blue shift of 0.85 nm. For instance, drop in the main positive band in the 90 fs – 2.5 ps time window is $\sim$45\% in V1 compared to a $\sim$28\% in V2. However, in V3 the signal is almost non-decaying throughout the detection axis. \si{Figure S7} shows that the differences in the decay dynamics of V1-V3 bands are significant beyond the error bars. Based on the above results, we can conclude that at least the amplitudes of exponential rates are excitation wavelength dependent, if not the rates themselves. Note that porphyrin nanotubes are known to form bundles of a few tens of nanotubes\cite{Kim2014,Leishman2015,Leishman2016a}. \si{Figure S8} shows that all the above 2DES features survive in case of bundles as well suggesting only weak inter-nanotube interactions with only a perturbative effect on the excitonic couplings within the nanotubes. The presence of early $T$ CPs and their rapid broadening prompted us for a deeper investigation of the CP$_L$ region using the polarization-controlled P-2DES experiments described below.\\

\subsection*{Selecting $Q_x-Q_y$ mixed states through polarization control}\label{polpop}

Following our earlier works\cite{bhattacharyya2023low, sahu2025isolating}, the transition dipole product for four light-matter interactions can be represented in a convenient notation. For example, an all \textit{x} or \textit{y} polarized transition dipole pathway, can be denoted as $\left\langle xxxx\right\rangle$ and $\left\langle yyyy\right\rangle$, respectively, with a suppressed notation for the electric field polarization. The all-parallel (PA) electric field polarization sequence is defined as  ($0^o,0^o,0^o,0^o$), the perpendicular pulse sequence PE as  ($90^o,90^o,0^o,0^o$), and the isotropic or magic angle (MA) sequence as  ($54.7^o,54.7^o,0^o,0^o$). With these definitions, orientational average between the lab and molecular coordinates yields $\left\langle xxxx\right\rangle_{PA}$ = 1/5 and $\left\langle xxxx\right\rangle_{PE}$ = 1/15. When mixed $Q_x-Q_y$ states are involved then only the cross terms such as $\left\langle xyxy\right\rangle$, $\left\langle xxyy\right\rangle$, etc. are specific to such mixing whereas isotropic terms such as $\left\langle xxxx\right\rangle$ and $\left\langle yyyy\right\rangle$ can arise without any mixing. The isotropic terms are eliminated by the $P$ polarization sequence defined as ($0^o,0^o,+60^o,-60^o$) and shown in \si{Figure~\ref{figure1}c}. The $P$ sequence can also be represented as $(PA-3PE)/4$ with $\left\langle xxxx\right\rangle_P = \left\langle yyyy\right\rangle_P = 0$, such that only the mixed $Q_x-Q_y$ states are selectively probed. This polarization sequence has been recently proposed by Zanni and co-workers\cite{farrell2022polarization} in the context of CP-selective 2D-IR spectroscopy and by us\cite{bhattacharyya2023low, sahu2025isolating} in the context of isolating vibronically enhanced anisotropic quantum beats from spectator vibrations. Our above argument shows that the $P$ sequence is sensitive to mixed $Q_x-Q_y$ states which need not arise exclusively in the CP region. Below we apply these ideas to investigate the overlapping vibrational and electronic bands of the porphyrin nanotube.\\

\begin{figure}
\centering
\includegraphics[width=4 in]{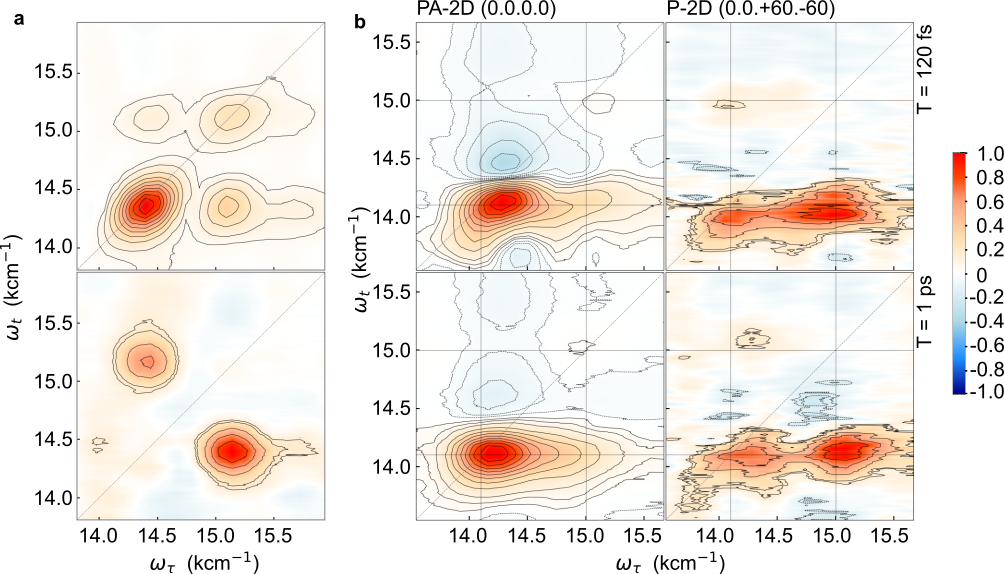}
    \caption{\footnotesize \textbf{Polarization-controlled 2D spectra with only $Q_x-Q_y$ paths selected.} \textbf{a} PA (top) and P-2D (bottom) spectra of H$_2$Pc at $T$ = 1 ps. The DP$_L$/CP$_L$ ratio changes from 2.9 to 0.1 in going from PA to P-2D. \textbf{b} PA and P-2D spectra of the porphyrin nanotube at $T$ = 120 fs (top) and 1 ps (bottom). Contours in PA-2D case are drawn at 2\%, 5\%, 10\% and then steps of 10\%. Contours in P-2D are drawn at 15\%, 20\% and then steps of 20\%. DP$_L$/CP$_L$ ratio changes from 3.2 to 0.8 in going from PA to P-2D. All 2D spectra are normalized relative to their own maxima.}\label{figure3}
\end{figure}

The polarization settings are described in detail in \si{Section S1.3}. For the 2DES experiments, we obtain polarization extinction coefficients of 1420 for the probe and 403 for the pump measured at the sample location. For the PP experiments, the pump extinction coefficient is 4498. We first tested our polarization settings on a D$_{2h}$ symmetry pthalocyanine molecule (29H,31H-phthalocyanine, H$_2$Pc), with non-degenerate $Q_x$ and $Q_y$ polarized transitions. \si{Figure~\ref{figure3}(left)} shows the observed suppression of the DP$_L$ relative to CP$_L$ due to elimination of isotropic $\left\langle xxxx\right\rangle_P$ type terms. The ratio of DP$_L$/CP$_L$ changes from 2.9 to 0.1 from PA to P-2D case, indicating a 29$\times$ suppression of the diagonal peak. \\


\FloatBarrier

\si{Figure~\ref{figure3} (right)} contrasts the PA-2D and P-2D spectra of the porphyrin nanotube. The SNR in the P-2D case, calculated using the time-domain data\cite{thomas2023rapid} is 3.1, compared to 10.6 in case of PA-2D. The PA-2D shows strong diagonal peak at both 120 fs and 1 ps which overwhelms the CP$_L$ region. In contrast, when mixed $Q_x-Q_y$ states are selected in P-2D, a major suppression of the DP$_L$ region is seen along with a distinct CP amplitude near the $Q$ band shoulder. The DP:CP intensity ratio for the PA and P-2D cases was quantified by integrating the DP$_L$ and CP$_L$ areas and taking their ratio. The ratio of the DP$_L$:CP$_L$ signal is 3.2 in PA versus 0.8 in P-2D, that is, $\sim$4$\times$ suppression of the diagonal peak. Note that the suppression of DP$_L$ and survival of CP$_L$ amplitude was confirmed across multiple nanotube sample preparations. \\

Taken together, the observation of $\sim$10$\times$ imbalanced CP$_L$ at early $T$, its rapid broadening, and the surviving $Q_x-Q_y$ amplitude in P-2D, together establishes the presence of strong $Q_x-Q_y$ mixing across the main $Q$ band and its shoulders. However, this observation is not in line with the predictions\cite{Stradomska2010b, Vlaming2009} from the existing vibronic exciton models for the porphyrin nanotube which predict almost no $Q_x-Q_y$ mixing between them. Our initial simulations of porphyrin nanotube linear spectra (\textit{vide infra}) were also consistent with these theoretical predictions. The question then arises as to what explains our experimental observations of strong $Q_x-Q_y$ mixing? We will return to this question during the discussion of vibronic exciton model for the porphyrin nanotube. \\


\begin{figure}[H]
\centering
\includegraphics[width=4 in]{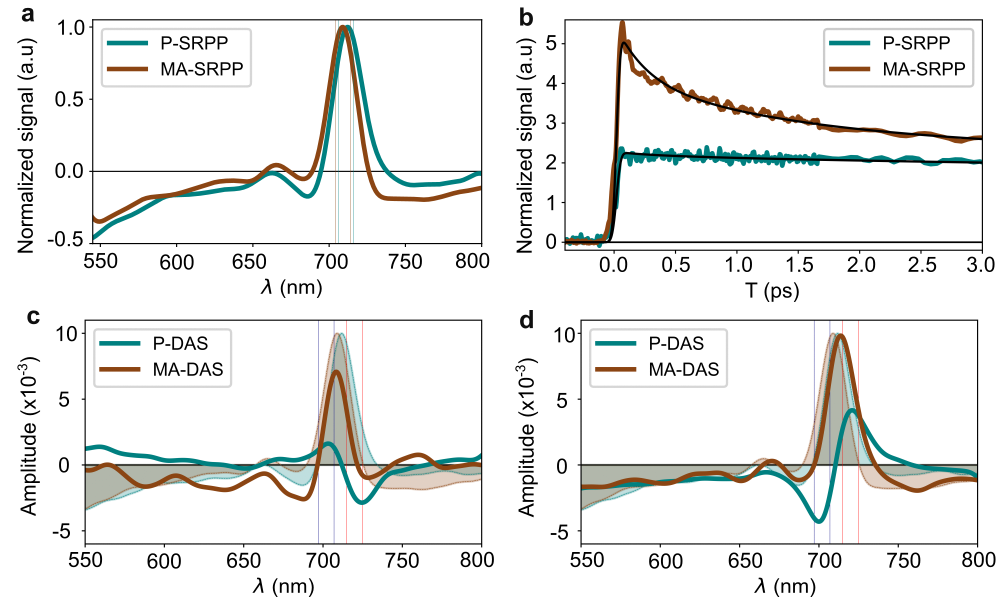}
    \caption{\footnotesize \textbf{Intra-$Q$ band internal conversion is revealed when only $Q_x-Q_y$ paths are selected.} \textbf{a} Normalized SRPP of nanotubes at $T$ = 1 ps measured for the MA versus $P$ experiments. The GSB/ESE band is marked by the vertical dashed lines. \textbf{b} GSB/ESE band integrated signal for the MA and P-SRPP experiments. The vertical dashed lines in the main positive band in panel \textbf{(a)} mark the respective bandwidths of integration. All the traces have been normalized at long $T$ in 200-220 ps range as shown in \si{Figure S11}. The P-SRPP signal is $\sim$0.45x weaker than the MA-SRPP signal at $T$ = 250 fs. \textbf{c} Global analysis of the SRPP data with the decay associated spectra (DAS) for MA and $P$ cases for the 238 fs time constant. \textbf{d} Global analysis of the SRPP data with DAS for MA and $P$ cases for the 1.34 ps time constant. The shaded traces in the background correspond to the scaled SRPP spectrum at $T$ = 1 ps. The integrated signal in the two 10 nm bands on the blue and red side of the GSB/ESE band is shown in \si{Figure S13}. }\label{figure4}
\end{figure}

Selection of mixed $Q_x-Q_y$ states also suggests that one could simply do a polarization-controlled spectrally-resolved pump-probe experiment\cite{farrell2022polarization} (P-SRPP) and still be able to selectively probe the mixed $Q_x-Q_y$ states. \si{Figure~\ref{figure4}a} compares the SRPP spectra recorded for the isotropic (MA) and $Q_x-Q_y$ specific ($P$) experiments at \textit{T} = 1 ps. The full SRPP data sets are shown in \si{Figure S9-S10}. The P-SRPP spectrum shows a $\sim$3 nm redshift in the main GSB/ESE band and a reduced (increased) ESA contribution to the red (blue), but with a similar overall shape as MA-SRPP. \si{Figure~\ref{figure4}b} compares the kinetics in the MA-SRPP and P-SRPP cases in the marked 10 nm GSB/ESE band. A markedly absent decay is seen in the P-SRPP trace where the signal primarily arises from the CP$_L$ region (\si{Figure~\ref{figure3}}). The observed lack of decay amplitude in the CP$_L$ region of the PE-2D spectra \si{Figure ~\ref{figure2}b} is consistent with the lack of decay amplitude seen in the integrated GSB/ESE band. \si{Figure S11} shows that when both the traces are normalized at longer $T$ (200 - 220 ps), the signal evolution after $\sim$ 30 ps is comparable between the MA and $P$ schemes. Not surprisingly this timescale is in approximate agreement with the timescale of polarization anisotropy decay in chlorosomes reported by Struve and co-workers\cite{savikhin1995ultrafast}. \\

\si{Figure~\ref{figure4}(c,d)} presents the global analysis and the comparison of early $T$ dynamics between the MA and P-SRPP signals. As described in \si{Section S3}, the MA signal has been fitted with a 5-exponential model with freely floated initial parameters. The resulting timescales of 238 fs, 1.34 ps, 12.94 ps, 122.23 ps and 3.37 ns are in close agreement with the reported timescales\cite{Kano2002}. The P-SRPP data was then globally fitted with the same time constants that were obtained from the MA fits. The full data sets are presented in \si{Figure S10}. Interestingly, compared to MA-DAS, the P-DAS of the fastest two time constants show a dispersive line shape in the GSB/ESE positive SRPP band. This striking change is easily visualized in \si{Figure S13} by  integrating the signal in a 10 nm bandwidth in the blue and red side of the GSB/ESE band, marked as blue and red lines in \si{Figure~\ref{figure4}c,d}. \\

For the case of 238 fs P-DAS, the dispersive lineshape denotes energetic relaxation of the stimulated emission (SE) signal within the main $Q$ band. Based on the separation between the positive and negative lobes of the dispersive P-DAS for the 238 fs time constant, we estimate the energetic separation to be $\sim$16 nm or 315 cm$^{-1}$. The separation is estimated from the 238 fs DAS by assuming two Gaussians with different amplitudes and positions. The analysis is shown in the \si{Figure S12}. The separation is lesser than the separation between the main $Q$ band (709 nm) and the first vibronic shoulder(669 nm) (40 nm or 840 cm$^{-1}$) and between the first (669 nm) and the second shoulder (637 nm) (32 nm or 750 cm$^{-1}$). Overlaying the 238 fs DAS with the $Q$ band absorption spectrum (\si{Figure S13}) suggests that both the states lie within the main $Q$ band with the acceptor level $\sim$315 cm$^{-1}$ below the main $Q$ band. It should be again emphasized that the above contrast between the MA and $P$ polarization experiments is made possible by selectively probing the mixed $Q_x-Q_y$ pathways.\\


As shown in \si{Figure~\ref{figure4}d}, the slower time constant in the P-SRPP experiments also shows a dispersive DAS lineshape that is again distinct from the MA experiment and oppositely signed compared to the 238 fs DAS (with a blue rise and a red fall). The corresponding full P-SRPP data is shown in \si{Figure S10} and suggests that the positive maxima of P-DAS is close to the emission peak (720 nm versus 717 nm, \si{Figure S13}). We therefore assign it to the decay of the ESE signal. The decay of the ESA band, which appears as a dip at 687 nm in the P-SRPP experiment is seen as a prominent negative lobe in the P-DAS in \si{Figure~\ref{figure4}d} (compare brown versus green curve in \si{Figure~\ref{figure4}a}). We therefore attribute the dispersive DAS lineshape of the 1.34 ps time constant to a decaying, blue-shifted ESA signal superimposed with a decaying red-shifted ESE signal. This decay likely corresponds to the decay of the population to the low oscillator strength/dark states below the main $Q$ band which then contribute to a $>$ 3 ns long-lived ESA signal (\si{Figure S9}). Dark band bottom states have been previously proposed to be responsible for faster fluorescence decay lifetimes\cite{akins1996fluorescence} in porphyrin aggregates. The rapid decay of ESE signal to the dark states with a long-lived ESA signal is akin to that reported\cite{Jansen2024} for the case of photosynthetic chlorosome nanotubes by Jansen and co-workers, although with $\sim$10$\times$ faster SE decay in case of chlorosomes. The low oscillator strength states below the $Q$ band caused by energetic disorder were proposed to play a key role in the photosynthetic design by prolonging the excited state population lifetime. Our observations suggest that similar principles may likely be operative in the case of porphyrin nanotubes as well. \\

\begin{figure}
\centering
\includegraphics[width=5 in]{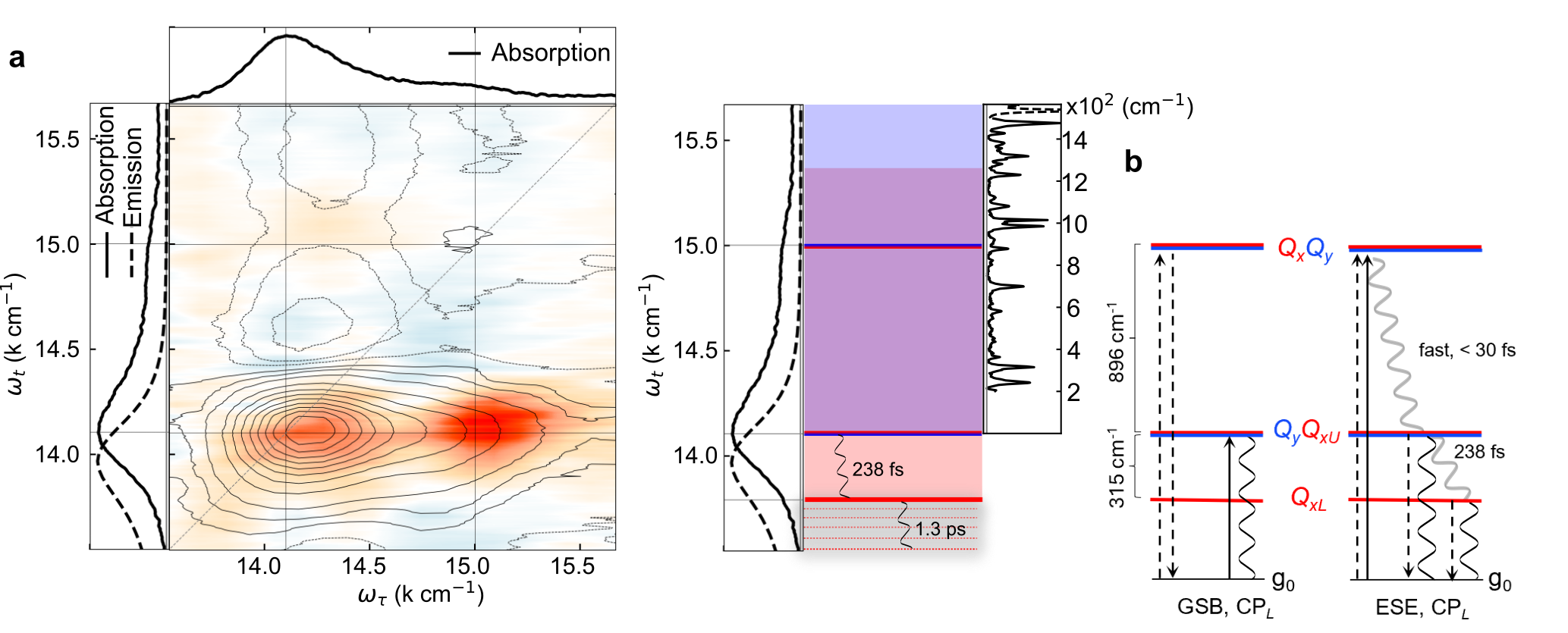}
    \caption{\footnotesize \textbf{Summary of mixed $Q_x-Q_y$ states deduced from PP and 2DES experiments.} \textbf{a.} (left) P-2D spectrum of porphyrin nanotubes at \textit{T} = 1 ps is overlaid with the PA-2D contour lines. The absorption and emission spectra are shown in the side panels. (right) Experimentally estimated positions of energy levels, their $Q_x-Q_y$ character and relaxation rates within the $Q$ band. The energy of the lowest horizontal line at 13.789 kcm$^{-1}$ is determined to be $\sim$315 cm$^{-1}$ below the main $Q$ band 14.104 kcm$^{-1}$ from the 238 fs dispersive P-DAS lineshape in \si{Figure~\ref{figure4}c}. The horizontal and vertical line at 15.0 kcm$^{-1}$ is determined by the location of the weak DP$_U$ peak in the PA-2D spectrum (\si{Figure~\ref{figure3}}). Red, blue and mixed red-blue denote the $Q_x$, $Q_y$ and mixed $Q_x-Q_y$ manifolds. The gray region at the bottom represent the lower-lying weak oscillator strength states. \textbf{b.} Wavemixing diagrams for GSB (left) and SE (right) signals that can contribute at the CP$_L$ location. The length of the first and last arrow in the wavemixing diagram determines the ($\omega_{\tau}$,$\omega_t$) position of the corresponding 2D signal. Squiggly line represents the radiated signal field. Time runs horizontally such that the interval between the second and third interaction is the waiting time $T$. After two pump interactions, the GSB and ESE pathways represent population density matrix elements $\rho_{g_0,g_0}$ and $\rho_{e,e}(T)$, respectively, where $e$ represents any given excited state which relaxes with $T$. The $T$ relaxation, say between $\rho_{e_2,e_2} \rightarrow \rho_{e_1,e_1}$, is depicted by slanted squiggly lines. The subscripts $L$ and $U$ on the $Q_x$ state represent lower and upper states, respectively. The experimentally determined $\sim$315 cm$^{-1}$ $Q_{xU}-Q_{xL}$ energy gap and the $\sim$896 cm$^{-1}$ gap between the main $Q$ band and the first shoulder are marked.}\label{figure5}
\end{figure}
\FloatBarrier

\si{Figure~\ref{figure5}} summarizes the picture that emerges from the kinetic analysis. \si{Figure~\ref{figure5}a}(left) shows the P-2D spectrum of the nanotubes overlaid with the contour lines corresponding to the $T$ = 1 ps PA-2D signal in \si{Figure~\ref{figure3}}. \si{Figure~\ref{figure5}a}(right) shows the estimated positions of the energy levels based on the PA and P-2D spectra, and the DAS lineshapes (\si{Figures~\ref{figure2}--\ref{figure4}}). The maximum amplitude in P-2D lies close to the first $Q$ band shoulder, seen as a weak DP$_U$ at 14.975 kcm$^{-1}$ ($\sim$668 nm) along the excitation axis and at the main $Q$ band position at 14.104 kcm$^{-1}$ (709 nm) along the detection axis. The mixed nature of the first $Q$ band shoulder is represented by a mixed red-blue horizontal line in \si{Figure~\ref{figure5}a}. \\


As we noted earlier, since the CP$_L$ signal in \si{Figure~\ref{figure2}} shows fast changes with $T$ -- $\sim$10$\times$ imbalanced CP$_L$ at 30 fs, its rapid broadening (\si{Figure~\ref{figure2}}) and population relaxation (\si{Figure~\ref{figure4}}) -- it is not a GSB signal. If CP$_L$ was due to a common ground state between the $Q_x-Q_y$ transitions shared across several molecules in the nanotube, CP$_L$ and CP$_U$ are expected to be of equal strength, given by a product of $|\mu_{Q_y}|^2|\mu_{Q_x}|^2$ transition dipoles. Rather, a $\sim$10$\times$ stronger CP$_L$ implies that the first shoulder is a mixed $Q_x-Q_y$ band that manifests as fast unresolved internal conversion between the first shoulder and the main $Q$ band. This is illustrated in the GSB and ESE signal pathways (wavemixing diagrams) in \si{Figure~\ref{figure5}b} which both contribute as incoherent population signals at the CP$_L$ location. In case of the ESE pathway, a fast unresolved energy transfer process explains the imbalanced CP$_L$ amplitude at 30 fs. A slower 238 fs relaxation of the ESE signal within the main $Q$ band, between the $Q_{xU}-Q_{xL}$ levels separated by 315 cm$^{-1}$, as estimated from the P-DAS in \si{Figure~\ref{figure4}c} is also shown. The picosecond decay of the ESE signal to the low oscillator strength dark states below the main $Q$ band is shown as well.

\subsection*{Anisotropic out-of-plane vibrational quantum beats revealed by polarization control}

In the context of unambiguously identifying promoter vibrational motions which mix electronic states and promote internal conversion versus spectator motions which do not, our recent works \cite{bhattacharyya2023low, sahu2025isolating} show that the $P$ polarization sequence can also eliminate isotropic vibrational quantum beats arising from spectator motions and isolate the anisotropic vibrational quantum beats which uniquely report the presence of vibronically mixed $Q_x-Q_y$ states. This is because the mixed $Q_x-Q_y$ states lead to signal pathways such as $\left\langle xxxx\right\rangle_P$ and $\left\langle xyxy\right\rangle_P$, where the former isotropic term leads to quantum beats that do not involve a change in electronic character and is eliminated, whereas anisotropic quantum beats from the latter term only arise in the presence of $Q_x-Q_y$ vibronic mixing and should survive. Anisotropic vibrational quantum beats arise from non-totally symmetric vibrations which tune the relative energy gaps and lead to vibronic coupling matrix elements that are akin to the Albrecht B-term\cite{Albrecht1970} in Raman scattering. \\


In the above context, it is interesting to note that a striking $\sim$30$\times$ intensity enhancement for the low-frequency Raman active vibrations has been reported\cite{akins1996fluorescence} upon formation of nanotube aggregates (see the resonance Raman spectra for the 241 and 317 cm$^{-1}$ peaks in Figure 5 of ref. \cite{akins1996fluorescence}). Similar modes have been implicated\cite{Kobayashi2002} in $B-Q$ Herzberg-Teller coupling. Whether these modes promote $Q_x-Q_y$ mixing is unequivocally reported by the presence (or absence) of anisotropic vibrational quantum beats in the $P$ polarization sequence.

\begin{figure}[H]
\centering
\includegraphics[width=5 in]{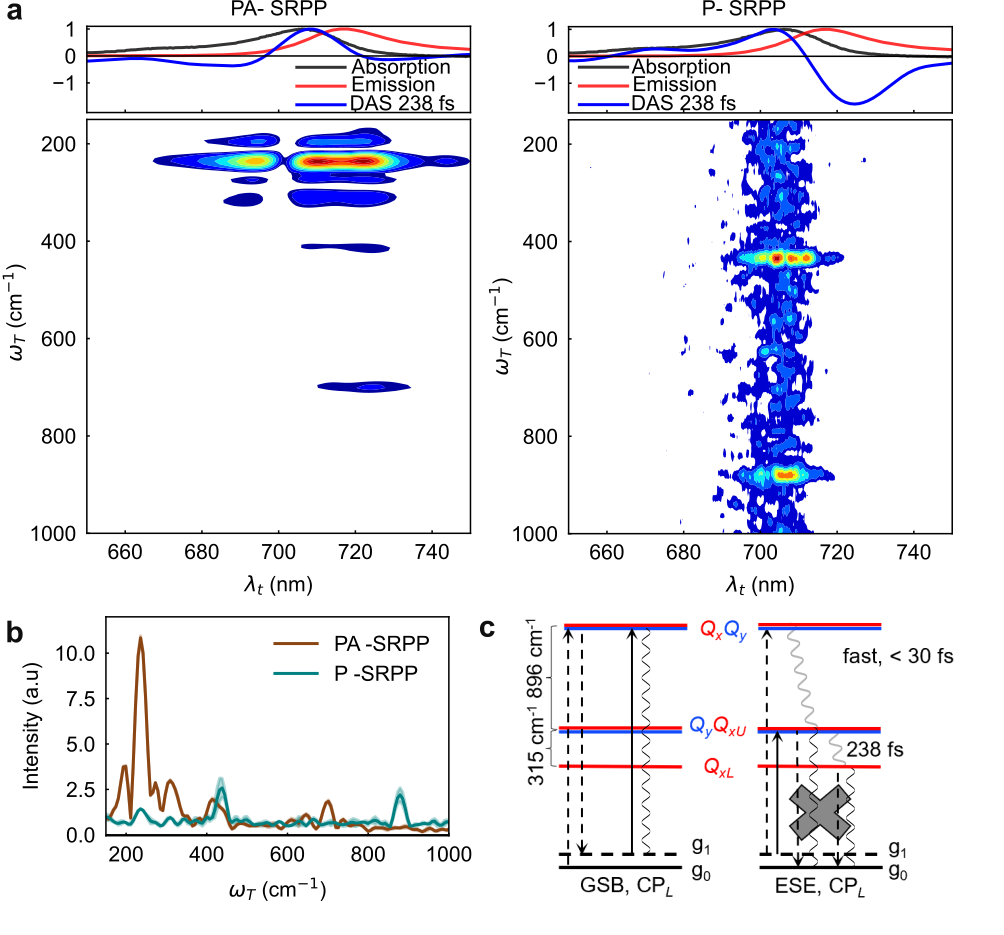}
    \caption{\footnotesize \textbf{Anisotropic quantum beats survive when only $Q_x-Q_y$ paths are selected.} \textbf{a} Quantum beat spectra from (left) PA-SRPP and (right) P-SRPP signal after removing the population decay. Both spectra have been individually normalized to the maximum signal with lowest contour of 15\% and 30\% for PA (left) and P (right) case, respectively. \textbf{b} Comparison of the beat amplitude across an integrated vertical band from 690-730 nm for the PA and P-SRPP data. The integrated signals are averaged across three datasets and plotted on the same scale without any multiplicative scaling of the collected signal. The corresponding errors bars are shown as bands. \textbf{c} Possible wavemixing pathways to describe the 880 cm$^{-1}$ vibrational quantum beats. The experimentally determined $\sim$315 cm$^{-1}$ $Q_{xU}-Q_{xL}$ energy gap and the $\sim$896 cm$^{-1}$ gap between the main $Q$ band and the first shoulder are marked. Slanted squiggly lines represent vibrational coherence transfer processes. Assuming such a process were feasible, the location of ESE pathways (length of the last interaction) will still be inconsistent with the observed spectral location of the quantum beats, under the main $Q$ band maxima along the detection wavelength axis. This is represented by a cross in the ESE pathway.}
    \label{figure6}
\end{figure}
\FloatBarrier

\si{Figure~\ref{figure6}a} shows the quantum beat amplitude as a function of detection wavelength obtained by removing the incoherent population background from the PA-SRPP (left) and P-SRPP (right) spectra. Strong beating amplitudes are immediately seen at 236 cm$^{-1}$ and 309 cm$^{-1}$, as may be expected from the already known 30x enhancement\cite{akins1996fluorescence} of Raman intensities for these vibrations. A comparison of quantum beat spectra between the PA- and P-SRPP experiments (\si{Figure~\ref{figure6}a,b}) reveals some surprising insights. When only the transitions involving mixed $Q_x-Q_y$ states are selected (P-SRPP), a remarkable suppression of the 236 cm$^{-1}$ and 309 cm$^{-1}$ beat is seen, immediately ruling out the role of these low-frequency ring deformation motions in $Q_x$-$Q_y$ mixing.  A quantum of excitation along the 317 cm$^{-1}$ deformation mode is near-resonant with the estimated $Q_{xU} - Q_{xL}$ energy gap (\si{Figure~\ref{figure5}b}). It may therefore be likely that the rapid $Q_{xU} \rightarrow Q_{xL}$ internal conversion is driven by low-frequency ring deformations which do not promote any $Q_x-Q_y$ mixing. \\

Curiously, the P-SRPP data shows two prominent quantum beats - 432 cm$^{-1}$ and 880 cm$^{-1}$ that survive the $P$ sequence. \si{Figure~\ref{figure6}b} plots the integrated quantum beat amplitude for the PA and $P$ experiments on the same scale. Both of the surviving quantum beats are obscured by the large amplitude lower frequency beats in the PA-SRPP case and appear as very weak peaks in the ground state Raman spectrum\cite{Akins1994}(\si{Figure S14}). Note that the elimination of the prominent 236 and 309 cm$^{-1}$ quantum beats and the survival of 432 and 880 cm$^{-1}$ beats was confirmed across multiple trials and shown as error bands in \si{Figure~\ref{figure6}b}. Both surviving quantum beats are known\cite{Akins1994} in the porphyrin nanotube and assigned to out-of-plane deformations of the pyrrole rings. Interestingly, both modes are not reported at higher pH where the porphyrin macrocycle is unprotonated, planar and of $D_{2h}$ symmetry (see Table 1 of ref.\cite{Akins1994}). Exactly similar trends - absent versus prominent vibrational modes are reported\cite{Waluk2017,Maes2011} for unprotonated and protonated porphin macrocycles, respectively, around 440-480 cm$^{-1}$ and 860-880 cm$^{-1}$. It is well known\cite{Fleischer1968} that upon protonation, the porphyrin ring undergoes a $D_{4h} \rightarrow D_{2d}$ ruffling deformation. Such a deformation has been implicated\cite{Waluk2017,Maes2011} in the activation of out-of-plane modes in protonated porphins observed in Raman\cite{Waluk2017} and fluorescence line-narrowing\cite{Maes2011} and phosphorescence\cite{Kruk2011} experiments. Resonance enhancement of out-of-plane vibrational modes upon excitation of out-of-plane electronic transitions has also been reported\cite{SpiroJACS1978} in strongly coupled $\mu$-oxo porphyrin dimers through the presence of $\alpha_{zz}$ component of Raman polarizability. The survival of anisotropic quantum beats from out-of-plane porphyrin deformation modes provides direct evidence for their participation in $Q_x-Q_y$ mixing and complements the previous reports on the activation of out-of-plane modes upon aggregation\cite{Akins1994}. \\

In order to decipher whether these surviving wavepackets are on the ground or excited electronic state, \si{Figure~\ref{figure6}c} analyzes the wavemixing signal pathways at the CP$_L$ location that are most consistent with the observation of 880 cm$^{-1}$ vibrational quantum beats. Recall that with the $P$ polarization sequence only the $\left\langle xxyy\right\rangle$, $\left\langle xyxy\right\rangle$ etc. pathways survive\cite{bhattacharyya2023low}. The left panel describes a density matrix element $\rho_{g_0,g_1}$, that is, a ground state vibrational coherence along the 880 cm$^{-1}$ surviving mode that proceeds through a mixed $Q_x-Q_y$ state on the first $Q$ band shoulder. The location of this signal pathway along the detection axis (length of the last interaction) is consistent with the appearance this quantum beat under the main $Q$ band. \si{Figure~\ref{figure6}c}(right) considers a similar pathway for the ESE signal with the initial excited state coherence $\rho_{Q_{xy0},Q_{xy1}}$. Such a pathway necessarily depends on the survival of vibrational wavepackets during internal conversion and their coherent transfer within the $Q$ band. This is typically associated with spectator vibrational motions\cite{Sahu2023, Jean1995} because they do not sample the mixed vibronic (anharmonic) coordinates and therefore are not susceptible to rapid dephasing. Even so, assuming secular Redfield approximation which maintains the frequency of transferred wavepackets, such an ESE pathway will contribute at the very tail of the $Q$ band ($\sim$880 cm$^{-1}$ below the main $Q$ band) making it inconsistent with the experimentally observed location under the main $Q$ band maxima. Based on the above arguments, we assign the surviving anisotropic quantum beats to ground state vibrational wavepackets corresponding to the out-of-plane porphyrin macrocycle deformations. Similar arguments are valid for the 432 cm$^{-1}$ vibration as well which we also assign to the ground electronic state. Given the $\sim$410 cm$^{-1}$ FWHM of the main $Q$ band, it is likely that the GSB pathway for the 432 cm$^{-1}$ vibrational quantum beats proceeds through vibronically mixed $Q_x-Q_y$ excited states that are within the main $Q$ band. The vibronic exciton model for the porphyrin nanotubes (\textit{vide infra}) confirms this argument.\\

It is quite interesting to note that a quanta of excitation along the 880 cm$^{-1}$ surviving mode is quite close in resonance with the energy gap between the main $Q$ band and the first shoulder that is estimated directly from the 2D spectra in \si{Figure~\ref{figure3}}. Jonas and co-workers have proposed\cite{tiwari2013electronic,Tiwari2017, Tiwari2018} a functional role for near-resonant vibrations that tune the donor-acceptor energy gap, in internal conversion through a non-adiabatic energy funnel. Our recent theoretical work predicts\cite{bhattacharyya2023low, sahu2023vibrations} that such vibrations are anisotropic and should survive the $P$ polarization sequence. Rapid internal conversion within the $Q$ band (\si{Figure~\ref{figure2}}), mixed $Q_x-Q_y$ states in the CP$_L$ region (\si{Figure~\ref{figure3}}), and prominent anisotropic vibrational quantum beats at near-resonant frequencies (\si{Figure~\ref{figure6}}), together provide \vt{the first direct} experimental evidence for functional vibronic couplings in large disordered photosynthetic aggregates at room temperature. \\

\subsection*{Disorder enhances $Q_x-Q_y$ vibronic mixing}

The above arguments still leave a central question unanswered as to what causes the rapid intra-$Q$ band internal conversion. We elaborate on this question below. Vibronic exciton models for cylindrically symmetric porphyrin nanotubes, including our own simulations discussed below, predict no $Q_x-Q_y$ electronic mixing -- main band is $Q_x$ polarized the first shoulder is $Q_y$ polarized. Theoretical work on vibronic dimers\cite{Tiwari2017,Tiwari2018} shows that the vibronic coupling matrix elements for near-resonant vibrations scale as $(\hat{q}_A - \hat{q}_B)\omega d \sin \theta$, where $\omega$ is the vibrational frequency, $d$ is the FC displacement, and $\hat{q}_{A,B}$ are vibrational coordinate operators on the respective site. Here $\hat{q}_A - \hat{q}_B$ is the energy gap tuning coordinate. $\theta$ is the electronic mixing angle given by $\frac{1}{2} \tan^{-1}(2J/\Delta)$, where $J$ is the Coulomb coupling and $\Delta$ is the site energy gap. Electronic mixing is a \textit{necessary} pre-requisite for vibronic mixing to occur within the overlapping vibrational-electronic bands. Therefore, the absence of $Q_x-Q_y$ electronic mixing precludes any role of $Q_x-Q_y$ vibronic mixing in promoting internal conversion. We therefore decided to revisit the vibronic exciton models for the $Q$ band of porphyrin nanotubes to understand whether the overlap between the dense vibrational-electronic bands of the aggregate is indeed inconsequential. 

\begin{figure}
\centering
\includegraphics[width=6 in]{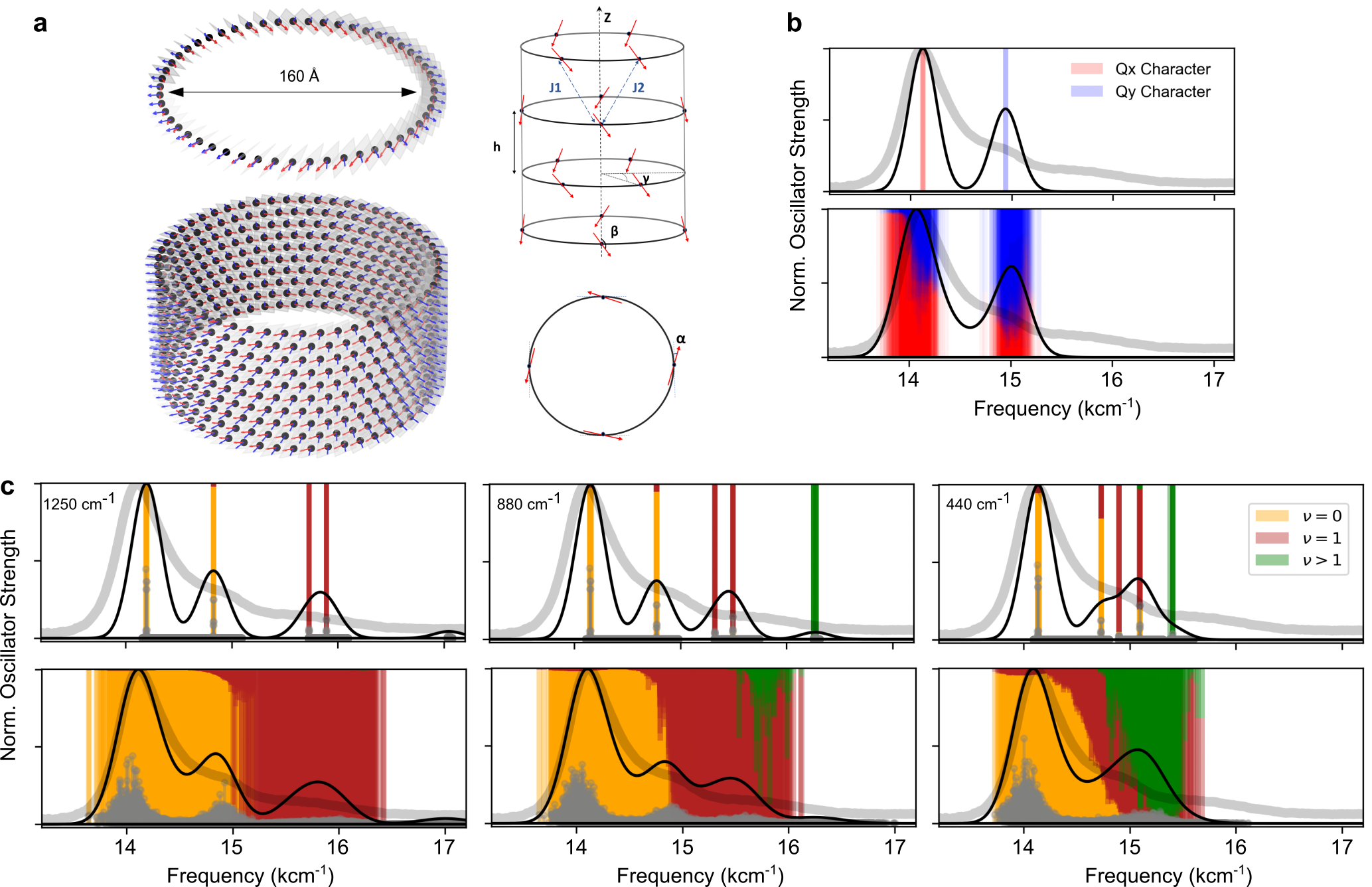}
    \caption{\footnotesize \textbf{Disorder enhances $Q_x-Q_y$ vibronic mixing.} \textbf{a.} Nanotube structural parameters. The experimentally estimated lattice spacing of $\sim$9.5 \r{A} and nanotube diameter of 160 \r{A} together lead to a ring of 53 molecules. Each ring is rotated by $\gamma =$ 3.4$^o$ (half index) and stacked to form a nanotube of 100 rings. Each site has Q$_{x,y}$ transition dipoles. Only the $Q_x$ dipole is shown in the right panel with its tangential and axial angle, $\alpha$ and $\beta$, respectively. \textbf{b.} Normalized oscillator strength (OS) spectra calculated from the electronic Hamiltonian (\si{Eqn.~S7}) without (top) and with (bottom) energetic disorder. The experimental absorption lineshape is shown in the background. The simulated lineshape is shown as thin black line. The electronic character of the states is plotted with OS cutoff of 0.05. \textbf{c.} Simulated OS spectra for the vibronic Hamiltonian with one explicit vibration in the system Hamiltonian with frequency (left) 1250 cm$^{-1}$, (middle) 880 cm$^{-1}$ and (right) 440 cm$^{-1}$. Top versus bottom rows show the calculation without versus with energetic disorder with OS sticks scaled by 250$\times$ and 5$\times$, respectively. The vibrational character is plotted with relative OS cutoff of 0.01. Full details of the calculation are presented in \si{Section S5}.}\label{figure7}
\end{figure}

We model the porphyrin nanotube with a vibronic exciton Hamiltonian with an explicit quantum vibration as part of the system Hamiltonian to fully capture the near-resonant non-adiabatic mixing between vibrational and electronic states. Only nearest neighbor electronic interactions are assumed. To restrict the basis sets size in case of the nanotubes with total 5300 molecules, we also assume one-particle approximation (1PA) which underestimates\cite{Sahu2020} resonant vibronic mixing effects by constraining ground state vibrational excitations to be only on the site of electronic excitations. The full details of the Hamiltonian and the experimentally constrained parameters are given in \si{Section S5}. Briefly, analytical ultracentrifugation experiments\cite{Russo2011}, and our own fluorescence(\si{Figure S16}) and pump-probe measurements (\si{Figure S15}) show that the diacid monomer (\si{Figure~\ref{figure1}a}) does not exist as a monomer at pH 4. Instead it should be a small aggregate (proposed\cite{Russo2011} as `ring') of $\sim$26 molecules. Thus, compared to previous works\cite{Stradomska2010b} which assume the `ring' aggregate to be the starting diacid monomer and float the nanotube structural parameters to obtain the best fit linear spectra, we adopt the following significantly more constrained approach. We first determine Huang-Rhys (HR) factors and transition dipole strengths by fitting the linear spectra for a known\cite{Russo2011} $H$ aggregate porphyrin dimer (\si{Figure S1, S19}). This is followed by floating the structural parameters to fit the experimentally obtained `ring' absorption spectrum (\si{Figure S1,S20}). These include the tangential and axial angles $\alpha$ and $\beta$ for $Q_x$ and $Q_y$ transition dipoles, respectively, as depicted in \si{Figure\ref{figure7}a}. The net $Q_x-Q_y$ angle is fixed at the experimentally determined\cite{Ogilvie2019} angle of 75$^o$. Having fixed the structural parameters, we then construct stacks of rings such that the nanotube diameter is experimentally constrained\cite{Russo2011} to be 160 \r{A} with previously estimated\cite{Stradomska2010b} center-to-center distance of 9.5 \r{A}. \\


\si{Figure~\ref{figure7}b} (top) shows the purely electronic spectrum of the nanotube where red (blue) denotes $Q_x$ ($Q_y$) character of the main $Q$ band and its first shoulder, respectively. The radial component of the $Q_y$ transition dipoles stack as $H$ aggregates within a ring. The long axis component of the $Q_x$ polarized dipoles across the stack of rings act like a $J$ aggregate whereas within a given ring they act like an $H$ aggregate. No $Q_x-Q_y$ mixing is evident from the electronic characters of the two bands. \\


Single-particle studies on chlorosome nanotubes\cite{Kohler2016} from green sulfur bacteria suggest that disorder in relative orientation and distance between the transition dipoles, transition energies and the nanotube radius is quite typical with ensemble averaged energetic disorder ranging upto $\sim$200 cm$^{-1}$ (see Section S6 of ref.\cite{Kohler2016}). Previous vibronic exciton models\cite{Stradomska2010b} of porphyrin nanotubes have incorporated an energy dependent Gaussian linewidth of $\sigma = $ 242--726 cm$^{-1}$ to phenomenologically model energetic disorder within the $Q$ band. Instead, we directly introduce energetic disorder in the electronic Hamiltonian (\si{Eqn.~S5}). The picture in \si{Figure \ref{figure7}b} becomes dramatically different even when a nominal value of Gaussian energetic disorder ($\sigma$ = 200 cm$^{-1}$) is introduced while maintaining the $Q_x-Q_y$ degeneracy on the individual sites. Electronic mixing is now seen across the entire $Q$ band with oscillator strength weighted $Q_x$ electronic character of 22\% and 49\% in the two main higher and lower energy regions of oscillator strength, respectively. The electronic character is quantified using \si{Eqn.~S7} and summarized in \si{Table S5}. Thus, disorder seems to be the vital ingredient that breaks electronic symmetry between the eigenstates to allow for $Q_x-Q_y$ electronic mixing, the prerequisite condition for intramolecular vibrations to contribute through vibronic mixing. \\

\FloatBarrier

The next question is whether near-resonant vibronic mixing that is well understood\cite{Jonas2018} in dimers, and now made possible in porphyrin nanotubes because of disorder induced $Q_x-Q_y$ electronic mixing, can survive in large disordered aggregates ? Our previous work\cite{Patra2022} on vibronic coupling in aggregates larger than dimers suggested that constructive interference effects between vibronic couplings scale up with aggregate size with vibronic splittings (and resonance widths) scaling as $N^2$. However, \textit{a priori} energetic disorder may be expected to suppress any such constructive interference\cite{Tiwari2018} effects in large aggregates. \si{Figure~\ref{figure7}c} calculates the oscillator strengths for the vibronic Hamiltonian for different known vibrational frequencies of the nanotube aggregate. Since the electronic character is now significantly more complicated than a dimer, we use the idea that resonant mixing allows for quantum admixtures of $\ket{v=0}-\ket{v=1}$, $\ket{v=1}-\ket{v=2}$, etc. states with each part of the mixture carrying a different electronic character. This allows us to ascertain the amount of $Q_x-Q_y$ vibronic mixing by tracking the oscillator strength weighted vibrational character of any given state. The vibrational character is quantified using \si{Eqn.~S8} and summarized in \si{Table S5}. A prominent\cite{Akins1994} high frequency C-C/C-N in-plane stretching mode at $\sim$1250 cm$^{-1}$ produces only negligible vibronic mixing (3\%) without or with disorder. In contrast, the out-of-plane mode frequency of 880 cm$^{-1}$ produces 5\% 0-1 mixing which enhances to 19\% in the presence of disorder. Notice the increased tilt of the 0-1 boundary. The 0-1 mixing increases to 20\% for the lower frequency 440 cm$^{-1}$ vibration and enhanced to 55\% in the presence of disorder. Now significant 0-1 mixing is seen within the main $Q$ band as well. It should be emphasized that the disorder induced enhancement of vibronic mixing suggested by the above models is expected to be increase\cite{Sahu2020} significantly when unrestricted vibrational basis sets (compared to 1PA), hot ground states and multiple vibrations\cite{Tiwari2018,bhattacharyya2023low} are treated explicitly.\\

The enhanced vibronic mixing range for the lower frequency vibration explains how the prominent off-resonant 432 cm$^{-1}$ anisotropic quantum beat can promote internal conversion even within the main $Q$ band. The simulation results on the porphyrin nanotubes are consistent with the predictions from previous dimer models\cite{Tiwari2018, bhattacharyya2023low} that energetic disorder can indeed enhance the width of resonant vibronic mixing. Note that we have also tested other aggregate geometries and the central conclusion about the role of disorder in promoting vibronic mixing is model agnostic (see \si{Section S5.5}). It is important to point out that in the above model we have assumed that all modes can tune relative donor-acceptor energy gap and therefore equally participate in vibronic mixing, which need not be true for all vibrational motions. The survival of only specific out-of-plane anisotropic vibrational modes does indeed point to this possibility. In this regard, detailed structural characterization of porphyrin nanotubes with cryo-electron microscopy, similar to that conducted\cite{Deshmukh2024} for cyanine nanotubes by Deshmukh et al., can provide the structural insights necessary for further refinement of the existing vibronic exciton models.\\

\section*{Conclusions} \label{section3}
To best of our knowledge this is the first instance where previously hidden details of $Q_x-Q_y$ vibronic coupling and fast intraband population relaxation within the $Q$ band of porphyrin nanotubes are revealed by a combination of polarization-controlled 2DES and pump-probe spectroscopy. Our experimental and theoretical observations strongly suggest that internal conversion within the overlapping vibrational-electronic bands of large disordered aggregates is driven by robust, resonance insensitive vibronic couplings which manifest as rapid loss of 2D frequency correlations and anisotropic vibrational quantum beats in the data. The rich dynamics arising from intraband $Q_x-Q_y$ mixing is likely overwhelmed in the isotropic experiments by the stronger and more number of transitions between $Q_x-Q_x$ and $Q_y-Q_y$ manifolds. Wavelength dependent relaxation\cite{JansenJPCB2023}, low-energy dark states\cite{Jansen2024} and rapid exciton delocalization\cite{Zigmantas2014} have all been reported for chlorosomes and similar observations in porphyrin nanotubes motivate artificial light harvesting templates that are based on chlorophyll-like chromophores, which may provide the right parameter regime for mimicking photosynthetic design principles. \\

Our findings about the survival of resonant vibronic mixing effects in large chlorosome-like aggregates at room temperature also carries direct implications for the active debate\cite{scholes2017using,Miller2020} on the functional role of vibronic mixing in driving photosynthetic energy transfer and complements the recent evidence\cite{Beck2022} for vibronic coherences in cyanobacterial protein supercomplexes from Beck and co-workers.  Enhanced delocalization\cite{Womick2011} of vibronic excitons resulting from near-resonant vibrational-electronic mixing in photosynthetic dimers has been reported at room temperature. The dual role of energetic disorder - first enabling electronic mixing, and then enhancing the energetic range of vibronic mixing across the entire $Q$ band - is quite striking and suggests that disorder may be a vital design principle for energy funneling in large and disordered photosynthetic aggregates. As opposed to the  localizing nature of disorder in quantum transport and non-radiative quenching of excitons\cite{RaoNature2024} by high-frequency vibrations, our observations suggest that vibronically enhanced quantum transport may be possible in a parameter regime where energetic disorder is of the order of dense low-frequency vibrations with weak reorganization energies. Disordered aggregates such as chlorosome nanotubes are predicted\cite{CaoPRL2016} to operate in the more robust two-dimensional limit of quantum diffusion where a fine balance of electronic couplings, dephasing, and energetic disorder is crucial to attain optimal transport. The survival of vibronic mixing driven internal conversion at room temperature adds yet another ingredient and motivates refined quantum transport models with a fully quantum vibrational bath, using for instance, effective modes\cite{Bittner2014} or a path-integral description\cite{Makri2022}, to adequately capture vibronic exciton delocalization beyond the Born-Oppenheimer approximation. \\

 \section*{Methods}\label{section4}

\subsection*{Sample Preparation}

{\footnotesize Here we briefly describe the procedure for the preparation of tubular aggregates of meso-tetra(4-sulfonatophenyl) porphyrin (TPPS). Further details about the purity of the isolated tubes and bundles are described in \si{Section S1.1}. TPPS was obtained from Santa Cruz Biotechnology as dihydrochloride salt. A dark green 200 $\mu$M TPPS stock solution is made in deionized water (DI), and the pH is adjusted to 3 by adding 1 N HCl. For nanotube preparation, this solution is then added to equal volume of 1.5 M HCl/DI solution in parts, which causes the colour to shift from dark green to light green. The resulting 100 $\mu$M solution is left undisturbed in the dark for multiple days after which the 490 nm absorption peak was seen to not rise further. This sample exhibits a clear absorption spectrum without contamination from the ring peak at 434 nm. The sample has a peak Q band OD of 0.405 at 709 nm measured in a 500 $\mu$m pathlength cuvette. }\\
	

\subsection*{Optical Setup}
{\footnotesize The detailed schematics of the partially collinear white-light 2DES and PP setups are reported elsewhere\cite{bhat2023,thomas2023rapid} and described in \si{Section S1.2}. Briefly we conducted impulsive pump-probe (PP) and 2DES experiments with a 100 kHz shot-to-shot detection approach. The input to the pump and probe arms consists of a YAG generated white light continuum (WLC). The pump and probe spectra at the sample location are overlaid with the $Q$ band of the porphyrin nanotube in \si{Figure~\ref{figure1}b}. In the 2DES experiment, the pair of phase-locked pump pulses are generated using a birefringent wedge based common path interferometer (CPI)\cite{Cerullo2014}. A combination of group delay dispersion compensated chirped mirrors and glass wedges are used for fine tuning the optical dispersion. The pump pulse duration of $\sim$10 fs, obtained after maximizing the two-photon interferometric autocorrelation generated by focusing into a SiC photodiode, is shown in \si{Figure S2a,b} and is sufficient to routinely excite vibrational coherences up to $\sim$1450 cm$^{-1}$. A global fit of the spectrally-resolved pump-probe (SRPP) data across the probe WLC bandwidth shows an instrument response function (IRF) full width at half maximum (FWHM) of $\sim$ 37 fs. The dispersion in signal rise time as a function of probe wavelength is only $\sim$ 27 fs over $\sim$ 200 nm bandwidth in the main signal region (\si{Figure S2c,d}). Field autocorrelation of the pump-pulse pair was routinely conducted to ensure the identical nature of the pump pulses -- real and absolute value Fourier transformed spectra showed deviations of $\sim$2\%, thus minimizing phasing artifacts that can arise from non-identical pump pulses. }\\

{\footnotesize In the partially collinear pump-probe geometry adopted in our setup, the pump and probe polarizations can be controlled independently. The optical layout of the polarizers and waveplates and the resulting characterization of polarization purity is described in detail in \si{Section S1.3}. At the sample location the polarization extinction ratios for the pump and probe pulses are 403 and 1420, respectively. The pump extinction ratio in the P-SRPP experiment is 4498. The measured probe power ratios through the analyzer for a 120$^\text{o}$ relative angle (+60$^o$ polarizer and -60$^o$ analyzer) were the same as the theoretical expectation of 1/4 to within the error bar of the measurement. A 100 kHz shot-to-shot rapid scan approach\cite{bhat2023,thomas2023rapid} has been implemented for data collection and described in \si{Section S1.4}. The sample \% transmission was confirmed to be linear across the range of pulse energies (\si{Figure S3}) used in the experiments.}\\

\subsection*{Vibronic Exciton Model For Simulations of Linear Spectra}
{\footnotesize Details of the vibronic exciton Hamiltonian and experimentally constrained model parameters for the H-dimer, ring and tube aggregate are provided in \si{Section S5}. Our vibronic exciton Hamiltonian, which has been detailed in previous works\cite{bhattacharyya2023low, Sahu2020} in the context of vibronic exciton dimers, treats one vibration explicitly in the system Hamiltonian to allow for numerically exact treatment of vibronic mixing along that vibrational coordinate without any Born-Oppenheimer approximation for the aggregate. A fully quantum treatment is necessary\cite{Sahu2020} to adequately describe near-resonant vibronic mixing effects such as resonance width, intensity borrowing and energy transfer rates. The effect of other prominent intramolecular Raman-active vibrations\cite{akins1996fluorescence} is modeled as Gaussian lineshapes with energy independent FWHM. Such a necessity arises because the exact basis set size increases exponentially as $N \times (n_{e,vib})^V \times( n_{g,vib})^{V(N-1)}$, where $N$ is the number of electronic states -- 2 per molecule in case of $Q_x, Q_y$ states, $n_{g(e),vib}$ denotes the total number of vibrational quanta on the ground (excited) electronic states and $V$ denotes the number of explicit vibrational modes in the system Hamiltonian.  For larger aggregates such as rings (number of sites $M$=26 and $N=2M$) we make the restricted 3-particle approximation (3PA, \si{Section S5.4}). For nanotubes with total 5300 molecules ($M$=100 stacked rings with 53 molecules/ring, \si{Section S5.5}), the basis set is significantly larger, and therefore we have to make the one-particle approximation (1PA) where the vibrational excitation and electronic excitation are constrained to be on the same site, that is, $n_{g,vib}$ = 1. Similar to previous works, simulation of the nanotube necessitates such approximations. Our previous work\cite{Sahu2020} has shown that when it comes to describing near-resonant vibronic mixing effects in the optical spectra, 1PA works much better in J aggregates compared to H aggregates. \si{Figure S17} compares the effect of 3PA versus 1PA on the ring aggregate where a noticeable change in oscillator strengths is seen within the first vibronic shoulder. The electronic couplings are estimated using the method of Knoester and co-workers\cite{Stradomska2010b} by assuming extended dipoles with $\pm$q charges separated by 5 \r{A}. The electronic and vibrational characters for the nanotubes are estimated using \si{Eqns. S7-S8}. The presence of vibronic mixing effects arising from low- or high-frequency vibrations is then ascertained by tracking the $Q_x-Q_y$ electronic character and the 0-1 vibronic character in the absorption sticks within the $Q$ band. Note that the constrained approach to fitting the linear spectra is significantly more constrained than previous approaches in that we do not attempt to obtain a dedicated\cite{Stradomska2010b} model for the nanotubes by freely floating structural parameters such as the diameter, number of molecules spanning the circumference, and axial and tangential angles, etc. and in treatment of energetic disorder which is included directly into the system Hamiltonian.}

 \section*{Data Availability}
 All data required for interpretation and verification of experimental results and reproducing the theoretical model are given in the paper and the Supplementary Information.
 
 \section*{Code Availability}
 Codes for the vibronic exciton models presented in the paper are available from the corresponding
 author upon reasonable request.

\bibliography{ArtificialAggsREFS}

\bibliographystyle{achemso}

\section*{Acknowledgements}
{\footnotesize A.S.T. acknowledges the Prime Ministers' Research Fellowship, MoE, India. V.N.B. acknowledges the DST-Inspire research scholarship. C.R. and I.R. acknowledge the research fellowship from IISc Bangalore. This work is supported in part by funding from Anusandhan National Research Foundation (No. CRG/2023/000327 and No. SCP/2022/000018) and Department of Biotechnology (No. BT/PR38464/BRB/10/1893/2020).}

\section*{Author Contributions}
{\footnotesize V.T. conceptualized the project. A.S.T. prepared all the nanotube samples and performed the 2DES, pump-probe and preliminary polarization-controlled experiments. C.R. performed the vibronic exciton model calculations. V.N.B. jointly led the effort to build and optimize the 2DES and pump-probe spectrometer along with A.S.T. V.N.B. and I.R. optimized and performed the polarization-controlled experiments. I.R. performed the initial vibronic exciton model calculations and analyzed the data from polarization-controlled experiments. V.T., A.S.T. and C.R. wrote the manuscript with inputs from all authors.}

\section*{Competing Interests}
{\footnotesize The authors declare no competing interests.}

\end{document}